\documentclass[nofootinbib,aps,prd,10pt,superscriptaddress,showpacs,preprintnumbers,eqsecnum]{revtex4-1}

\usepackage{amsmath}
\usepackage{dsfont}
\usepackage{multirow}
\usepackage{graphicx}
\usepackage{mathrsfs}

\usepackage{amsfonts}
\usepackage{amssymb}
\usepackage[dvips]{color}

\begin{document}
\newcommand\be{\begin{equation}}
\newcommand\ee{\end{equation}}
\newcommand\bea{\begin{eqnarray}}
\newcommand\eea{\end{eqnarray}}
\newcommand\bseq{\begin{subequations}} 
\newcommand\eseq{\end{subequations}}
\newcommand\bcas{\begin{cases}}
\newcommand\ecas{\end{cases}}
\newcommand{\p}{\partial}
\newcommand{\f}{\frac}

\title{BKL maps and Poincar\'e sections}

\author{Orchidea Maria Lecian}
\email{lecian@icra.it}
\affiliation{Sapienza University of Rome, Physics Department, Piazzale Aldo Moro, 5 - 00185 Rome, Italy}


\begin{abstract}
Cosmological billiards arise as a map of the solution to the Einstein equations, when the most general symmetry of the metric tensor is implemented, under the BKL (named after Belinski, Khalatnikov and Lifshitz) paradigm, for which points are spatially decoupled in the asymptotical limit close to the cosmological singularity.\\
Cosmological billiards in $4=3+1$ dimensions for the case of pure gravity are analyzed for those features, for which the content of Weyl reflections in the BKL maps requires definition of a $3$-dimensional restricted phase space. The role of Poincar\'e sections in these processes is outlined. The quantum regime is investigated within this framework: as a result, $1$-epoch BKL eras are found to be the most probable configuration at which the wavefunctions have to be evaluated; furthermore, BKL eras containing $n>>1$ epochs are shown to be a less probable configuration for the wavefunctions.\\
This description of the dynamics allows one to gain information about the connections between the statistical characterization of the maps which imply the different symmetry-quotienting mechanisms and the characterization of the semiclassical limit of the wavefunctions for the classical trajectories, for which the phenomenon of 'scars' on the wavefunction is found for other kinds of billiards.

\end{abstract}


\maketitle
\section{Introduction\label{section1}}
Cosmological billiards are a very powerful description of the solution of the Einstein equations in the case of the most general symmetry hypothesized for the metric tensor, as shown in \cite{KB1969a} \cite{Khalatnikov:1969eg}, \cite{BK1970}, \cite{BLK1971}, \cite{Lifshitz:1963ps} \cite{LLK}, where the one-dimensional BKL map (named after Belinski, Khalatnikov and Lifshitz) was established. In particular, under the BKL hypothesis, space gradients are neglected with respect to time derivatives in the asymptotic limit close to the cosmological singularity. Here, the solution to the Einstein field equations is mapped to the behavior of a billiard ball in a closed domain, which undergoes elastic bounces on the sides of the billiard table, and experiences free-flight (according to the geodesics induced by the metric) evolution \cite{misn}, \cite{Misner:1969hg}, \cite{chi1972}, \cite{Misner:1994ge}. In \cite{Chernoff:1983zz}, \cite{sinai83} and \cite{sinai85}, \cite{Kirillov:1996rd}, the geometrical interpretation of the statistical analysis was achieved by the definition of $2$-variable maps.\\
The interest in this description of the cosmological singularity has been renewed by the discovery of new structures \cite{Damour:2001sa}, \cite{Damour:2000hv}, \cite{Damour:2002fz}, \cite{Damour:2002cu}, \cite{Damour:2002et}, \cite{hps2009}, \cite{bel2009}.\\
\\
In $4=3+1$ dimensions, two kinds of billiard are obtained: the 'big billiard', for which pure gravity is taken into account, and for which the original BKL scheme is recovered, and the 'small billiard', which coincides with $1/6$ of the big billiard, and which is defined by the symmetry constraints of the theory \cite{Damour:2010sz}.
In this paper, the attention will be devoted to the case of $4=3+1$ pure gravity.\\
In this case, the corresponding cosmological billiard is triangular, and the dynamics of the billiard ball that describes the solution to the Einstein equations under the BKL hypothesis is fully understood by means suitable symmetry-quotienting mechanisms, which allow one to schematize the evolution of the billiard ball by means of the so called BKL maps. These maps can be defined for the reduced phase space \cite{Damour:2010sz} or for the Upper Poincar\'e Half Plane (UPHP) \cite{Balazs:1986uj}, \cite{yellow}. The choice of the map depends on the content of Weyl reflections \cite{terras} in the trajectory of the billiard ball \cite{Lecian:2013cxa}. In fact, the presence of Weyl reflections is not explicitly outlined in the maps for the reduced phase space, and, as a result, different subcases have to be considered. These maps, and their content of Weyl reflections, permute the role of the Kasner exponents, which represent the projective directions in the asymptotical limit close to the cosmological singularity.\\
In particular, in this paper, the cases will be analyzed, for which the investigation on the UPHP with a $3$-dimensional invariant measure has to be followed \cite{corn1982} \cite{Damour:2010sz}.\\
\\
The quantum version of the model can be envisaged according to the choice of boundary conditions that can be imposed on the wavefunction, i.e. Dirichlet \cite{csordas1991}, \cite{Benini:2006xu}, \cite{Kleinschmidt:2009cv}, \cite{Kleinschmidt:2009hv} and Neumann \cite{Forte:2008jr}, \cite{Graham:1990jd} boundary conditions 
\cite{graham1991}
, to the symmetries that characterize the geometrical features of the billiard table \cite{Fleig:2011mu}
 and of the dynamics \cite{Kleinschmidt:2010bk}, \cite{Graham:1994dr}, \cite{Lecian:2013cxa}, and to the information that has to be looked for \cite{Graham:1990jd}, \cite{mk11}.\\
\\ 
One of the greatest achievements of BKL maps is to enclose the continuous dynamics of the billiard chaotic motion into a discrete map, which is statistically interpreted as the Poincar\'e return map on the sides of the billiard (i.e. the sides of the billiard are considered as a 'special' Poincar\'e section), and on which the geometrical features of the billiard table as well as the properties of the dynamics allow to implement several symmetry-quotienting mechanisms.\\
From the quantum point of view, on the other hand, the sides of the billiard are imposed boundary conditions, according to the quantization procedures. The $3$-dimensional energy-shell reduced Liouville measure allows one to consider Poincar\'e sections different form the sides of the billiard. The knowledge of the complete dynamics and of the symmetry-quotienting mechanisms allows one to extract physical information from these different Poincar\'e sections.\\
According to the statistical features of cosmological billiards, the most probable configurations are given by particular periodic phenomena on the UPHP. i.e. $1$ epoch BKL eras are the most probable ones. Because these periodic configurations on the UPHP do not correspond to periodic configurations inside the Unit Hyperboloid (UH) \cite{Damour:2010sz}, when the unquotiented map is considered, it is necessary to implement the appropriate tools on the UPHP. As a result, the configurations corresponding to the classical $1$-epoch BKL eras are discovered to be probabilistically favored also in the quantum regime.\\
The wavefunction corresponding to the semiclassical limit for the quantum version of several kinds of billiards is found to be characterized by the appearance of 'scars', which are structures described by the enhancement of the absolute value of the wavefunction in correspondence of short-period periodic orbits. For the case of cosmological billiards, the consideration of Poincar\'{e} surfaces of section not coinciding with any sides of the billiard table allows one to gain information about the properties acquired by the wavefunction when short-period periodic orbits only are considered, and to connect these properties with those of the statistical description of the discrete dynamics obtained by means of the different symmetry-quotienting mechanisms.\\ 
\\
As a result, rigorous connections are found between $(i)$ the role of Poincar\'{e} surfaces of section, $(ii)$ the properties of the statistical maps which define the different symmetry-quotienting mechanisms, $(iii)$ the semiclassical limit of the wavefunctions with respect to the classical trajectory on which it is evaluated, $(iv)$ the definition of the quantum probabilities for the semiclassical limit of the wavefunctions, $(v)$ the characterization of the wavefunctions according to the different angular velocities at which the billiard ball crosses the surface of section in the classical regime, $(vi)$ the regions of the restricted phase space containing particular periodic configurations corresponding to the appearance of structures called 'scars', observed for billiards defined in a different geometry, but for which no description was available for cosmological billiards.\\ 
\\
The paper is organized as follows.\\
In Section \ref{section1}, the framework within which cosmological billiards emerge is depicted.\\
In Section \ref{section2}, the main features of cosmological billiards are recalled.\\
In Section \ref{section3}, the $3$-dimensional Liouville measure for special Hamiltonian systems is applied to some problems in cosmological billiards, where the content of Weyl reflections on the maps on the UPHP requires the consideration of a Poincar\'e section different form the sides of cosmological billiards.\\
In Section \ref{section4}, the implementation of the quantum regime, and, in particular, of the quantum BKL maps is faced by means of the tools acquired in the previous sections.
In section \ref{section5a}, the role of short eras is examined, both at the classical level and in the quantum regime, according to the features of the phase space. The possible connections between some kinds of periodic trajectories and the phenomenon of scars for the wavefunctions in the semiclassical regime are linked by the statistical characterization of the dynamics obtained by the implementation of different symmetry-quotienting mechanisms.\\
In Section \ref{section5}, the difference between the geometrical features of the billiard table and the properties of the dynamics is stressed within the framework of the analysis developed.
Brief concluding remarks end the paper.\\ 
\section{Basic Statements\label{section2}}
\begin{figure*}[htbp]
\begin{center}
\includegraphics[width=0.7\textwidth]{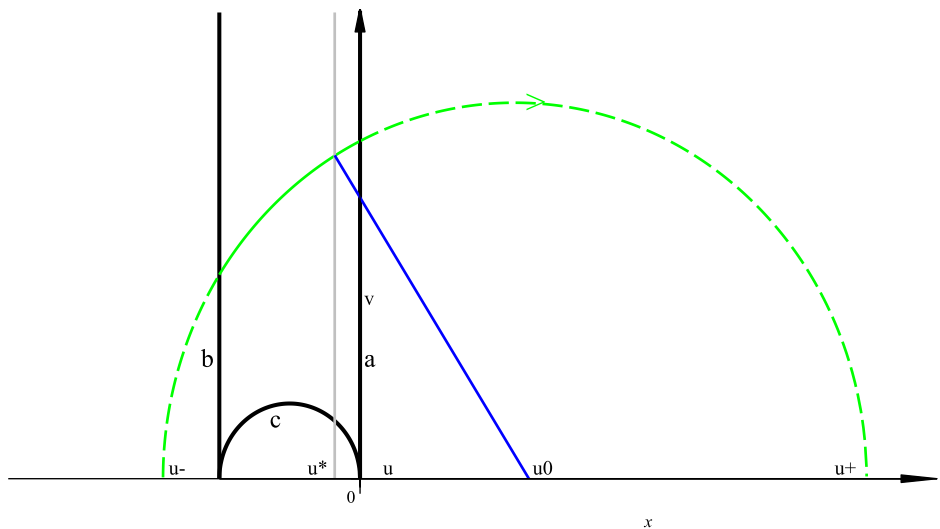}
\caption{\label{figura1} An oriented geodesics on the UPHP is plotted in green (light gray): the billiard trajectories are parameterized according to the oriented endpoints $u^+$ and $u^-$; the radius of the corresponding half-circle is plotted in blue (dark gray), and the center $u_0$ is indicated. A Poincar\'e section $u=u^*$ is plotted in gray.}
\end{center}
\end{figure*}

In the case of a general inhomogeneous space-time, close to the cosmological singularity \cite{llft} \cite{Misner:1974qy}, 
the Iwasawa decomposition of the metric tensor reads
\begin{equation}\label{iwa}
g_{ij}(x^0,x^i)=\sum_{a=1}^d e^{-2\beta^a(x^0,x^i)}N^a_{\ \ i}(x^0,x^i)N^a_{\ \ j}(x^0,x^i).
\end{equation}
The relevant degrees of freedom are those expressed by the variables $\beta$, while the other degrees of freedom, encoded in the lapse function $N$ and in the shift vector $N^a$ are considered as 'frozen'.\\
As a result, the dynamics of the early universe is described by the chaotic motion of a massless billiard ball in a triangular potential well delimited by infinite potential walls, and moving along the geodesics of the metric 
\begin{equation}\label{betametric}
d\sigma^2=\sum_{a,b=1}^{d}G_{ab}d\beta^a d\beta^b=\sum_a\left(d\beta^a\right)^2-\left(\sum_ad\beta^a\right)^2
\end{equation}
According to this description, the bounces against the potential walls are elastic, and the geodesic free-flight evolution between each bounce is defined by the Kasner velocities $v^a$, defined by the Kasner exponents $p^a$ as $p^a\equiv v^a/\sum_av^a$, which obey the constraints $\sum_ap^a=(\sum_ap^a)^1=1$.\\
In this framework, the cosmological singularity, located at $t\rightarrow0$ in Eq. (\ref{iwa}), corresponds to the time variable $\tau\rightarrow\infty$, where $d\tau=-dt/\sqrt{g}$. 
The metric (\ref{betametric}) defines a Unit Hyperboloid (UH), inside which the chaotic motion along the geodesics takes place. This geodesic motion can be further parameterized according to a radial coordinate $\rho$, and the remaining degrees of freedom are those encoded by the angular $\gamma^a$ variables, which define the surface of the UH, such that $\beta^a\equiv\rho\gamma^a$, $\gamma_a\gamma^a=-1$.\\
The solution of the Hamiltonian constraint\footnote{For a thorough review of the solution of the Hamiltonian constraint according to the Misner-Chitr\'e variables, see \cite{Montani:2007vu}} with respect to the variable $\lambda=\ln\rho$ is equivalent to the projection of the geodesic motion onto the surface of the UH. By geometrical transformations, it is possible to project the $\gamma$ surface on the unit disk, and then on the UPHP. In $4=3+1$ dimensions, when pure gravity is considered, the cosmological billiard table is triangular, as illustrated in Figure \ref{figura1}, where the sides of the billiard table $a$, $b$, $c$ are drawn. They are defined by the equations $a: u=-1$, $b: u=0$, $c: u^2-u+v^2=0$, where the variables $u$ and $v$ parameterize the UPHP.\\
On the UPHP, geodesics are (generalized) half-circles, parametrized by their oriented endpoints. For a $2$-dimensional system, the complete phase space is $4$-dimensional, but, given the symmetries of the problem and the elastic bounces of the billiard walls, it is possible to show that the complete phase space is $2$-dimensional, given by the oriented endpoints of the geodesics.\\
The $3$ corners of the billiard are due to the directions toward which the $3$ anisotropic scale factors reach $\infty$.\\
For the geometry of the billiard, the succession of trajectories of the billiard ball can be ordered according to the corner they take place in, and to the orientation (i.e. according to the oriented endpoints). A trajectory joining any two walls is named a (Kasner) epoch, while a set of trajectories taking place inside the same corner of the billiard is named a (Kasner) era. As there $3$ corners and two possible orientation for each corner (the orientation being defined by the first epoch of each era), there are $6$ kinds of (Kasner) eras. As geodesics are oriented, epochs are named after the corner where they take place, and according to the sides of the billiard they fly within: in Figure \ref{figura1}, a $ba$ epoch is plotted. Accordingly, eras are named after the first epoch they contain: i.e., a set of $n$ epochs taking place within the two sides $a$ and $b$ is called an era of the $ba$ type is the first epoch is an oriented trajectory starting from the side $b$ and ending on the side $a$.\\
On the UPHP, the geodesics are circles described by the equation
\begin{equation}\label{2a}
v^2+(u-u_0)^2=r^2,
\end{equation}
where the center $u_0$ and the radius $r$ depend on the oriented endpoints $u^+$ and $u^-$ as
\begin{subequations}\label{3}
\begin{align}
&u_0=\tfrac{u^++u^-}{2}\ \ \ \  r=\tfrac{u^+-u^-}{2}\label{3a}\\ 
&u=u_0+r\cos\theta\ \ \ \ v=r\sin\theta \label{3b}
\end{align}
\end{subequations}
In Figure \ref{figura1}, a trajectory of the billiard is sketched, and its oriented endpoints on the $u$ axis of the UPHP are indicated.\\
The reduced phase space $(u^-,u^+)$ is coordinatized by the oriented endpoints, and consists of several squared 'boxes' built of squared 'sub-boxes', which express the length of eras according to the variable $u^+$. The dynamics of the 'big billiard' is therefore described as a 'hopscotch game' on the 'hopscotch court' defined as the reduced phase space consisting of the two variables $u^-$ and $u^+$,where (Kasner) epochs and (Kasner) eras are defined as points on the 'hopscotch court'. Even though $6$ kinds of eras are possible, it is possible to chose a preferred orientation, such as the $ba$ one, onto which all the kinds of eras can be mapped, according to different symmetry-quotiented mechanisms that exchange the role of the Kasner parameters.\\
\\
\subsection{BKL maps\label{subsectionmap}}
The role of BKL maps is to define different symmetry-quotienting mechanisms for the Kasner Parameters. For the anisotropy of space postulated within this framework, indeed, the $3$ Kasner parameters are related to the $3$ space directions. According to the possibility of describing the dynamics on the UPHP by means of the reduced phase space, the BKL maps can be stated for the variables $(u^-,u^+)$: in the reduced phase space, the BKL maps act diagonally on the two variables. In \cite{Damour:2010sz}, the relation between the symmetry-quotiented maps and the complete maps are analyzed both for the single-variable version of the transformations, as well as for the double-variable version. In \cite{Lecian:2013cxa}, the features of these maps on the UPHP is examined, and the role of Weyl reflections is outlined.\\
All the maps are defined by identifying each era with an $n$-epoch era of the $ba$ type by identifing the first epochs of the era. In the $(u^-,u^+)$ reduced phase space, the first epoch of each era is represented as being parametrized by the oriented endpoints $u^+$ and $u^-$ in a squared domain ('box') delimited by $-2\le u^-\le -1$ and $n-1\le u^+\le n$ (for which the choice of (\ref{3a})) and (\ref{3b}) is consistent, as one can verify in Fig. \ref{figura1}.\\ 
\\
\begin{table}
\begin{center}
    \begin{tabular}{ | l | l | }
    \hline
    $ {\rm epoch\ \ map} $ & $ {\rm era\ \ map} $\\ \hline
    $ u^\pm\rightarrow \mathcal{T}_{ba} u^\pm=u\pm-1$ & $u_F^\pm\rightarrow\mbox{\large{$\mathsf{T}$}} u_F^\pm=+\frac{1}{u_F^\pm-[u_F^+]}-1$\\
    $ z\rightarrow T^{-1}z $ & $z\rightarrow\mathbf{T}z=\equiv T^{-1}SR_1T^{-n+1}z$\\ \hline 
    \end{tabular}
\end{center}
\caption{\label{table1} Quotiented maps for the billiard dynamics.\\
These maps act on points defining the epochs and eras of the $ba$ type, for which the variables of the restricted phase space have the ranges $-\infty\le u^-\le-1$, $u^+\ge 0$ for the epoch map, and $-2\le u^-\le-1$, $u^+\ge 0$ for the era map. The maps for the variables $(u^-,u^+)$ variables of the restricted phase space are given in \cite{Damour:2010sz}. The $1$-variable BKL map is recovered by marginalizing the variable $u^-$.\\
The maps for the complex variable $z=u+iv$ are given in \cite{Lecian:2013cxa}, by making use of the transformations $T(z)=z+1$,
$S(z)=-\tfrac{1}{z}$, and $R_1(z)=-\bar{z}$. The $1$-variable BKL map is obtained by evaluating the variable $z$ at $v=0$, and then by considering the oriented endpoint of the corresponding geodesics.}
\end{table}
The two-variable epoch map and the two-variable era map are recalled in Table \ref{table1} both for the restricted phase space $(u^-,u^+)$ and for the UPHP.\\ 
The epoch maps act on the oriented endpoints of the trajectories defining the free-flight evolution of the billiard ball and define the change to the new epoch by identifying the two sides of the billiard table which form the corner where the epoch takes place.\\
The era maps describe the succession of eras by describing the number of epoch in each era via the continued-fraction decomposition of the values of the variables $(u^-,u^+)$ of the first epoch of each era. It identifies the different sides of the billiard table when a new era starts.\\
All these maps generalize the original BKL map, which is a one-variable map for the auxiliary variable $u$, identified with $u^+$ in the statistical description.\\
According to these maps, for the one-variable map, the continued fraction decomposition of $u^+$ contains the succession of BKL eras in the dynamics, and $u^-$ can be interpreted as an initial condition, while, for the two-variable map, the variable $u^-$ encodes the 'past' succession BKL eras in the dynamics, according to the action of the map $\mbox{\large{$\mathsf{T}$}}$ in Table \ref{table1}.\\
The $1$-variable BKL epoch map and the BKL $1$ variable BKL era-transition map are obtained $i$) by marginalizing the variable $u^-$, such that the BKL $u_{>0}$ map is recovered for the variable $u\equiv u^+$; $ii$) by evaluating the geodesics on the absolute of the UPHP (i.e. for $v=0$), and then by considering the 'rightmost' endpoint , which corresponds to $u^+$, as depicted in Figure \ref{figura1}.
\\
Going the other way round, it is possible to 'tile' the UPHP by curvilinear (partially overlapping) domains, where epoch of the $ba$ type only are defined, according to the number $n$, as defined in \cite{Lecian:2013cxa}.\\
\paragraph{BKL Probabilities} 
Within this picture, probability $P_n$ of an era to contain $n$ epochs is given by the area of the pertinent subregion of the reduced phase space according to the invariant measure $\tfrac{1}{2}\tfrac{du^+du^-}{(u^+-u^-)^2}$ obtained from the reduced form $\omega(u^-,u^+)$
\begin{equation}\label{symp}
\omega(u^-,u^+)=\tfrac{1}{2}\ln\tfrac{du^-\wedge du^+}{(u^+-u^-)^2}.
\end{equation}
and reads
\begin{equation}\label{5}
P_n=\tfrac{1}{\ln 2}\ln\tfrac{(n+1)^2}{n(n+2)}.
\end{equation}
From the probability $P_n$ in Eq. (\ref{5}), $P_n$ is a positive monotonically decreasing function of $n$, such that $1$-epoch eras are the most probable. Furthermore, as shown on \cite{Damour:2010sz}, the dynamics of the billiard exhibits preferred directions, such that $n$-epoch eras are more probable when $n$ is odd. Combining the two information together, one can infer that the most probable configuration, also for the unquotiented big billiard, is one given by the periodic succession of $1$-epoch eras, characterized by the values $u^-=-\Phi$ and $u^+=\phi$, where $\Phi$ is 'Golden Ratio', $\Phi\sim 1,618$, and the 'Small Golden Ratio' $\phi$ is defined as $\phi+1=\Phi$. Moreover, because the BKL map is unstable under small perturbations, According to the study of the Ljapunov exponents of the problem, the BKL map for the variable $u^+$ is unstable under small perturbations, while the variable $u^-$ is not: a sequence of several $1$-epoch eras can be obtained by a small perturbation of the periodic configuration but this sequence is a finite one, i.e., after it, an era with $n\neq1$ epochs will take place sooner or later.\\
\\
\section{Invariant measures and Poincar\'e sections\label{section3}}
In this section, the definition of invariant measures for this model will be revised from \cite{Damour:2010sz}, its connection with the definition of a Poincar\'e section for the c Hamiltonian problems will be outlined, and the corresponding definition for the chaotic motion inside the unit hyperboloid UH will be proposed.\\
\\
\subsection{Energy-shell reduced Liouville measure}
The BKL probability for an $n$-epoch era to take place obtained by integrating the pertinent subdomain of the reduced phase space is originated by the analysis of the invariant measures for Hamiltonian systems. In particular, the probability $P_n$ in Eq. (\ref{5}) holds when a Poincar\'e return map for the dynamics is considered.\\
Nonetheless, when the complete geodesic flow is taken into account, i.e. when also the geodesic path is considered, the complete phase space is $3$-dimensional \cite{Damour:2010sz}. 
This way, the $(u^-,u^+)$ space is sheltered by a 'slab' of varying thickness $\mathcal{S}$, with $0\le \mathcal{S}\le \sigma(u^+,u^-)$. In other words, the energy-shell reduced Liouville measure $\Omega^{(3)}_L$ equals the product of the symplectic measure(\ref{symp}) by $d\mathcal{S}=d\theta/\sin\theta$, with $0\le\theta\le\pi$, such that
\begin{equation}\label{omega3}
\Omega^{(3)}_L=\omega(u^-,u^+)\wedge d\mathcal{S},
\end{equation}
and the billiard motion becomes a so-called
'special flow' \cite{corn1982}, i.e. a combination of uniform
motion in the ‘vertical’ $\mathcal{S}$ direction, , with discrete 'jumps' in the variables $u^+$ and $u^-$. The action of the unquotiented big billiard maps on the energy-shell reduced Liouville measure for the unprojected motion inside the UH has been examined in \cite{Damour:2010sz}.\\
\\
Indeed, the path along the geodesic is parameterized by the angle $\theta$; alternatively, one can choose to parameterize it according to the value $u^*$, where $-1<u^*<0$, i.e. the projection of the position of the billiard ball on to $u$ axis of the UPHP, such that $\cos\theta=(u_0-u^*)/r$. This way, the thickness of the slab defined in \cite{Damour:2010sz} reads
\begin{equation}\label{6}
\sigma(u^+,u^-)=\sigma_0(u^+,u^-)+\tfrac{1}{2}\ln\tfrac{1-\tfrac{\mid u_0-u^*\mid}{r}}{1+\tfrac{\mid u_0-u^*\mid}{r}}.
\end{equation}
\paragraph{Weyl reflections}
As analyzed in \cite{Lecian:2013cxa}, the BKL maps defined on the reduced phase space do not encode entirely the properties of the transformations, for which Weyl reflections are required. As a direct example, one sees that the value $\cos\theta=\mid u_0-u^*\mid /r$ implied in the definition of $\sigma(u^+,u^-)$ in Eq. (\ref{6}) changes its sign under Weyl reflections on the UPHP (as oriented endpoints are considered for the BKL map).\\
In the following, the energy-shell reduced Liouville measure will be a fundamental tool to implement the BKL maps on the reduced phase space by taking into account the Weyl reflections imposed on the UPHP. As a result, it will be possible to consider the symmetry-quotienting mechanisms outlined by the BKL maps for the Kasner coefficients, and to keep track of the Weyl reflections without mismatches.\\
Furthermore, the definition of BKL maps as in Subsection \ref{subsectionmap} allows one develop calculations for the sub-boxes of the reduced phase space, where these maps are defined, thus defining clearly the initial conditions and eliminating ambiguities.\\
\\
\subsection{Poincar\'e sections}
The role of Poincar\'e sections for the description of hyperbolic billiards will be explained in the following paragraphs. Poincar\'e sections can be considered when a Poincar\'e return map has to be established, and the physical content of the Poincar\'e return maps is better encoded in surfaces different from the sides of the billiard itself.\\
Poincar\'e sections will be considered on the UPHP and on the UH.\\
\\ 
\paragraph{Poincar\'e Sections on the UPHP}
It is easily verified that it is not possible to choose any section for the big billiard, different from the sides of the billiards, according to which the exact evolution of the big billiard is reproduced. Nevertheless, two strategies are possible: either one $i$) restricts the ranges of the variables $u^-$ and $u^+$ such that they parameterize only one kind of epochs. In the (preferred) case of the $ba$ type, this is achieved by imposing $-2\le u^-\le -1$ and $0\le u^+\le\infty$, or, equivalently, $ii$)on the UPHP, selects only the geodesics which belong to the tiling that classifies epochs of the $ba$ type. The details of these domain are recalled in Table \ref{table2}.\\
The vertical $\mathcal{S}(u^*)$ can be interpreted as a particular Poincar\'e section, through which the geodesics pass, different from the sides of the billiard, and through which a Poincar\'e return map can be searched. An example of a Poincar\'e section for the return map on the $b$ side of the billiard is given in Figure \ref{figura1}.\\
More in general, one  seen that the shape of the Poincar\'e section $\mathcal{S}$ is that of a geodesic (i.e. a generalized circle on the UPHP defined by $u=u^*$) that separates the $ca$ corner of the billiard for the other $2$ by exactly cutting the $ba$ corner. A Poincar\'e section of this kind is sketched in Figure \ref{figura1} by the vertical $u=const$ line. This way, one can deduce that any geodesics having this behavior, ore even more general ones, can be chosen for these purposes. More in detail, any geodesics $u=u^*$ with $-1\le u^*\le0$, or any half-circle intersecting two sides of the billiard
 can be considered a Poincar\'e section for the Hamiltonian flow. Going the other way round, one can interpret the sides of the billiard as special Poincar\'e sections (at which the billiard ball bounces) where the return map is considered.\\
\\
\paragraph{Poincar\'e Sections for the UH}
The geodesic chaotic motion inside the UH is governed by the metric (\ref{betametric}), decomposed along the radial direction $\rho$ and the angular variable $\gamma$, and the (gravitational) walls can be interpreted as well as the 'special' (i.e. complete) Poincar\'e sections where the Poincar\'e return map is considered. Projecting the dynamics on the $\gamma$ surface eliminates one degree of freedom for the model.\\
Nevertheless, it is worth remarking that one could think of performing the Poincar\'e returm map for the dynamics also inside the $\beta$ hyperboloid by choosing an appropriate Poincar\'e section for this case. Even though the shape of the UH can be 'visualized' in a Euclidean space, even though the (gravitational) walls can be represented as planes intersecting the UH in a Euclidean space, the geodesic path is given by the metric (\ref{betametric}). This way, and by comparison with the considerations for the UPHP of the previous Section, one sees that an appropriate Poincar\'e section inside the UH is any (according to the metric (\ref{betametric})) plane intersecting the light cone where the UH is placed. One easily sees that all such planes will have the same role as $\mathcal{S}(u^*)$.\\
In particular, such planes will isolate a corner of the billiard with respect to the other two, and ,in general, it is possible that such planes will not bisect any corner of the gravitational walls.\\
\\
\paragraph{Poincar\'e Sections and BKL maps}
It is straightforward to verify that the implementation of the BKL symmetries of the dynamics inside the UH, where the Hamiltonian constraint has not been solved yet, play the same role of the solution of the Hamiltonian constraint with respect to the variable $\lambda$ (which eliminates one degree of freedom by projecting the dynamics on the $\gamma$ surface). This would not be true if the BKL symmetries were not implemented, because of the 'anisotropic' behavior the complete big billiard exhibits with respect to the existence of a preferred direction in the succession of corners.\\
\\
As an example, one can choose a generalized (according to the metric (\ref{betametric})) plane, different form the gravitational walls, inside the light cone, of equation $\beta^a=const$. The chaotic motion of the point particle inside the UH is such that the particle will bounce against the $\gamma$ surface. Some trajectories will pass through the section $\beta^a=const$ and some will not. Anyhow, it is possible to identify all these trajectories by identifying the Kasner coefficients (by means of suitable Kasner transformations). Because the Kasner coefficients describe the asymptotic (toward the singularity) behaviors of the anisotropic scale factors, the Poincar\'e return map of the (now oriented) geodesics on  the $2$-dimensional Poincar\'e section ($\beta^a=const$ in this example) constitute a $2$-dimensional statistical description inside the UH for the solution of the Einstein equations.\\
For this representation of the solution of the Einstein equations, the phase space is $3$-dimensional: it consists of the two coordinates on the $\beta^a=const$ plane, sheltered by a 'slab' which accounts for the direction with which the trajectory pass through the generalized plane.\\
\\ 
\section{Quantum regime\label{section4}}
The quantum implementation of the model is obtained by solving the Hamiltonian constraint, such that a proper WDW equation is obtained, and then by considering the quantum properties of the corresponding scheme. The quantum regime of this model has been examined both for the UH variables (decomposed as $\beta^a=\rho\gamma^a$), and on the UPHP (for the variables $u$ and $v$).\\
For the UPHP, the wavefunction $\Psi$, defined for the variable $z=u+iv$ of the UPHP, obeys the eigenvalue equation
\begin{equation}\label{eigen}
-\Delta_{\rm LB}\Psi(z)=E\Psi(z),
\end{equation}
where $\Delta_{\rm LB}$ is the Laplace-Beltrami operator on the UPHP, $\Psi$ is decomposed along its modes for the variable $v$, and along its Fourier modes for the variable $u$, such that $\Psi=\sum_s\Phi_s$, where $\Phi_s$ is defined as
\begin{equation}\label{7a}
\Phi_s(u,v)=\sum_\mu c_\mu y^{1/2}\mathcal{K}_{s-\tfrac{1}{2}}(2\pi\mid \mu\mid v)exp[2\pi i\mu u]:
\end{equation}
in Eq. (\ref{7a}), the functions $\mathcal{K}$ are the modified Bessel-$K$ functions of the second kind. As a result, the wavefunction $\Psi$ explores the energy levels $s$ of the eigenvalue equation (\ref{eigen}), and scans the quantum BKL numbers $n$ through the decomposition along the $u$ direction.\\
In \cite{Lecian:2013cxa}, the complete wavefunction inside the $\beta$ hyperboloid has been taken onto account; the complete eigenvalue equation has been solved by requiring that the wavefunction factorizes into the $\rho$ and the $\gamma$ variables. As a result, the wavefunction and its $\rho$ derivatives have been shown to vanish in the vicinity of the cosmological singularity.\\
The boundary conditions for this functions have been examined according to different viewpoints. In particular, Neumann and Dirichlet boundary conditions have been chosen in order to stress on different properties.\\
Furthermore, in \cite{Lecian:2013cxa}, the symmetry which have to be taken onto account if one wants to interpret the full billiard wavefunction as a suitable junction of the wavefunctions of the small billiard have also been considered, according to both the symmetry of the billiard table and to the symmetries that govern the dynamics.\\
\\
\subsection{Quantum BKL configurations}
\begin{figure*}[htbp]
\begin{center}
\includegraphics[width=0.7\textwidth]{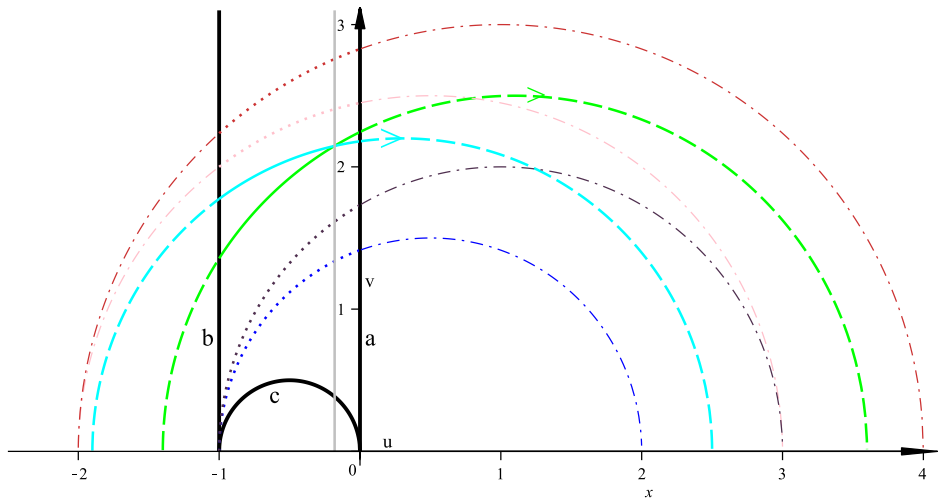}
\caption{\label{figura2} Two oriented geodesics are plotted on the UPHP: the green (gray) trajectory corresponds to a $4$-epoch BKL era, while the light blue (light gray) trajectory corresponds to a $3$-epoch BKL era. A Poincar\'e section $u=u^*$, where the two trajectories cross, is plotted in gray. The subdomains that tile the billiard table according to the content of BKL epochs in each era are sketched by the dotted lines, which are the geodesics $\mathcal{G}^M(4)$, $\mathcal{G}^m(4)$, $\mathcal{G}^M(3)$ and $\mathcal{G}^m(3)$, whose definition is recalled in Table \ref{table2}.}
\end{center}
\end{figure*}

Quantum BKL probabilites have been proposed in \cite{Lecian:2013cxa}, where they have been defined as the integral of the squared modulus of the wavefunction $\Psi$ on the pertinent regions of the UPHP, where the geodesics describing an $n$-epoch BKL era tile the billiard table. Nevertheless, this 'tiling' is not uniform, and these regions overlap. This problem can be faced in the reduced phase space $(u^-,u^+)$, where the subregions defining $n$-epoch BKL eras are squared 'boxes' and do not overlap. For this purpose, however, one has to consider the full $(u^-,u^+,u^*)$ description. Indeed, the Fourier decomposition along the $u$ direction of $\Phi_s$ is defined for $-1\le u \le0$, but detecting the mechanism of quantum BKL probabilities implemented by the Poincar\'e return map on the sides of the billiard can be made difficult by the choice of particular boundary conditions for the wavefunction.\\
In the following, this inconvenience will be removed by considering the full $(u^-,u^+,u^*)$ description, i.e. by considering a suitable Poincar\'e section defined by the generalized geodesics $u=u^*$, with $u^*$ not coinciding with any side of the billiard. Even though many configurations can be hypothesized, the use of BKL maps recalled in Subsection \ref{subsectionmap} will allow one to implement each problem for the intervals $-2\le u^-\le -1$, $u^+>0$ on the restricted phase space, and  $v>0$, $-1\le u\le0$ on the UPHP.\\
The oriented geodesics that define the trajectories of the billiard ball intersect $u=u^*$ are parametrized along each geodesic path by the variables
\begin{subequations}\label{8}
\begin{align}
&u^*\\
&v^2+(u^*-u_0)^2=r^2:
\end{align}
\end{subequations}
there are only $2$ equations for the set of three variables $(u^-,u^+,u^*)$, and this is required for the definition of the position of the billiard ball along a given geodesics if the variable $u$ has to be decomposed according to its Fourier modes in the definition of Maass wavefunctions.\\
As a result, the eigenfunctions for this model will read
\begin{equation}\label{maass}
\Phi_s(u,v)=(r^2-(u^*-u_0)^2)^{1/4}\sum_\mu c_\mu exp[\pi i\mu u^*]\mathcal{K}_{s-\tfrac{1}{2}}(2\pi\mid \mu\mid (\sqrt{r^2-(u^*-u_0)^2})),
\end{equation}
\\
As outlined in \cite{Lecian:2013cxa}, on the UPHP, the billiard table is tiled according to the number $n$ of epochs contained in BKL eras, where the tiles are delimited by the geodesics $\mathcal{G}^M(n)$ and $\mathcal{G}^m(n)$, whose radius and centers are recalled in Table \ref{table2}. Nevertheless, this tiling is not uniform, and the corresponding subregions partially overlap, as shown in Figure \ref{figura2}.\\
\begin{table}
\begin{center}
    \begin{tabular}{| l | l | l | }
    \hline
    $r_m(n)$ & $r_M(n)$ & $u_0(n)$ \\ \hline
    $n/2$ & $(n+2)/2$ & $(n-2)/2$ \\ \hline
    $n/2$ & $(n+2)/2$ & $(n-2)/2$  \\ \hline
    
    \end{tabular}
\end{center}
\caption{\label{table2} The values of the maximum radius $r_M(n)$ and of the minimum one $r_m(n)$, and of the center $u_0(n)$, specified for the value $n$, of the geodesics delimiting the curvilinear domains that tile the UPHP according to the presence of BKL epochs of the $ba$ type. Particular kinds of these domains are plotted in Figure \ref{figura2}.}
\end{table}
For this, given two different geodesics parameterizing an $n$-epoch era and an $n+1$-epoch era, respectively, it is possible to find a configuration for which they intersect at a given $-1\le u^*\le 0$ and a given $v$. The difference in the number of BKL epochs of the era they represent is given by the direction they pass the Poincar\'e section $u=u^*$, i.e. by the angle $\theta$, in Eq. (\ref{6}). This information is not entirely encoded in the oriented endpoints of the geodesics $u^-$ and $u^+$.\\
When the quantum wavefunction is considered, two different wavefunctions $\Phi_{s,n}$ and $\Phi_{s,n+1}$ acquire the same value, and so do their derivatives with respect to $v$: the only difference is given by the quantum BKL numbers $n$ and $n+1$, and this information is encoded in the derivatives with respect to $u^+$ and $u^-$.\\
\\
To investigate the physical information enclosed in the wavefunction $\Phi_s$, it is convenient to define the object $\Delta^{\pm\pm}_{n,n'}(u^-, u^+, {u^-}', {u^+}')$ as the ratio between two different derivatives of the wavefunctions, evaluated at the same point $u=u^+$ and $v=v(u^*)$ (through which an infinite number of geodesics pass, with different endpoints), which encode different successions of BKL epochs and eras, as  
\begin{equation}\label{delta}
\Delta^{\pm\pm}_{n,n'}(u^-, u^+, {u^-}', {u^+}')\equiv\frac{\left(\tfrac{d\Phi_s}{du^\pm}\right)_{u^\pm}}{\left(\tfrac{d\Phi_s}{du^\pm}\right)_{{u^\pm}'}}:
\end{equation}
when evaluated at two different endpoints $u^\pm$ and ${u^\pm}'$, it encodes the different successions of BKL epochs and eras, as for their continued-fraction decomposition and for the action of the map \mbox{\large{$\mathsf{T}$}}.\\
The interpretation of the one dimensional BKL map, i.e. that where only the  value $u^+$ is taken into account, consists in considering the values of $u^-$ as 'initial conditions'.
\\
For the case ${u^+}'=u^++1$, for large values of $n$, one obtains
\begin{equation}\label{a}
\Delta^{++}_{n,n+1}\simeq1+\tfrac{1}{n}
\end{equation}
where $m$ is the value of epochs contained in the \textit{previous} BKL era, according to the definition of the map $\mbox{\large{$\mathsf{T}$}}$.   
Eq. (\ref{a}) has the same asymptotical (for large $n$) series expansion of the ratio between the two BKL probabilities Eq. (\ref{5}) evaluated at $n$\ and at $n+1$, i.e. $P_{n}/P_{n+1}\simeq 1 + \mathcal{O}(1/n)$. This allows one to infer that the ratio between the two derivatives, evaluated at the same point $u^*$, and with the same boundary and initial conditions, states the asymptotical probability to find an $n$-epoch BKL era according to the same classical probability law (\ref{5}), for which BKL eras containing a number $n>>1$ of BKL epochs are probabilistically disfavored, and, in particular, $n$-epoch eras are more probable than $n+1$-epoch eras.\\
For any value of $n$ and $n'$, one has
\begin{equation}\label{b}
\Delta^{--}_{n,n+1}=\tfrac{u^+-u^*}{{u^+}'-u^*}=\tfrac{n-1+x_+-u^*}{n'-1+x_+'-u^*}:
\end{equation}
this ratio is extremized for small values of $n$, i.e. for $n=1$, at a fixed $n'\neq n$. In Eq. (\ref{b}), $x_+$ is the fractional part of $u^+$.\\
\\
By the action of the map $\mbox{\large{$\mathsf{T}$}}$ on $u^-$, one then learns that the most probable configuration, also form the quantum point of view, is one for which $u^-$ is close to the value of $\Phi$.\\
\\
As a result, it is possible to conclude that, also from the quantum point of view, the most probable configuration at which the wavefunction $\Phi_s$ has to be evaluated is (at least) close to that corresponding to the periodic configuration of $1$-epoch eras.\\  
\\
The point in having introduced $u^*$ is avoiding to establish any fixed boundary condition (such as Neumann or Dirichlet boundary conditions) on the wavefunction. In fact, it is in principle always possible to define a Poincar\'e section on the billiard table, where the wavefunctions do not vanish identically.\\
\\
\subsection{Quantum BKL probabilities}
Collecting all the ingredients together, it is possible to express the quantum BKL probabilities by means of a suitable integral on the pertinent regions of the reduced phase space. It is interesting to recall that it is not possible to define a Poincar\'e section \textit{a priori} on the restricted phase space $(u^-,u^+)$, because the energy-shell reduced Liouville measure (\ref{omega3}) is composed by the two elements $\omega(u^-,u^+)$ and $d\mathcal{S}$ as 'complementary' (or orthogonal).\\
The formal expression for the quantum BKL probability $\textbf{P}_{n,s}$ is then given by 
\begin{equation}\label{12}
{\bf P}_{n,s}=\int_{-2}^{-1}du^-\int_{n-1}^{n}du^+\int_{\theta_2}^{\theta_1}\tfrac{d\theta}{\sin\theta}\tfrac{1}{(u^+-u^-)^2}\mid\Phi_{s}(u^-,u^+,u^*)\mid^2,
\end{equation}
where the integration is performed according to the classical energy-shell reduced Liouville measure (\ref{omega3}) recalled in Section \ref{section3}. It is important to remark that here the BKL era map is being used, such that the orientation of the angle $\theta$ is always the same, and no sign ambiguities arise. Integrating the geodesic path $d\mathcal{S}(u^*)$ along a complete BKL (oriented) trajectory consists in considering the two values $u^*=-1$ and $u^*=0$, for which $\cos\theta_1=\mid u_0+1\mid$ and $\cos\theta_2=\mid u_0\mid$ (as one can easily verify from Figure \ref{figura1}).\\
With respect to the formal result obtained in \cite{Lecian:2013cxa}, where the integration was performed on the UPHP on overlapping domains, this way the problem is avoided by following the geodesic path inside the billiard and then by integrating the remaining functions on the reduced phase space.\\
Eq. (\ref{12}) defines the quantum BKL probabilities for a BKL era to consists of $n$ BKL epochs form the quantum point of view, and, in particular, the probability for a wavefunction to be evaluated on the first epoch of an $n$-epoch era.\\
\\
It is possible to express also the probability for an epoch to be the $n'$-th epoch of an $N'>n'$-epoch BKL era, as
\begin{equation}\label{13}
{\bf P}_{n',s}=\int_{-2-n'}^{-1-n'}du^-\int_{N'-n'}^{N'-n'-1}du^+\int_{\theta_2}^{\theta_1}\tfrac{d\theta}{\sin\theta}\tfrac{1}{(u^+-u^-)^2}\mid\Phi_{s}(u^-,u^+,u^*)\mid^2,
\end{equation}
where the quotiented $\mathcal{T}_{ba}$ map has been used for the evaluation of the integration boundaries, and the notation of \cite{Lecian:2013cxa} has been kept for the labellings of epochs.
\\
\subsection{The role of Poincar\'e surfaces of section\label{subsection4c}}
As already recalled, the original BKL map consists in labelling the bounces of the billiard ball on the surface of section constituted by the generalized geodesics $u=u^*$ according to the decreasing values of the variable $u^+$, and links the description of the continuous billiard dynamics to that of a discrete system.\\
Considering a different Poincar\'e surface of section allows therefore one to grasp information about the continuous dynamics without losing information about the BKL statistics.\\
From a very general point of view, this can be useful both at the classical level and in the quantum regime.\\
From the classical point of view, this procedure can help one gain information about the features of the dynamics if the motion on the billiard table is hypothesized for some reasons to be not a geodesic motion any more, but to still follow the geodesic trajectory, as, for example, if any kind of friction is present on the billiard table.\\
From the quantum point of view, even if the semiclassical limit of the wavefunction is obtained by evaluating the wavefunction on the classical trajectory, the features of the quantum models are such that the evolution of a wavefunction depends on the point at which the wavefunction is evaluated even though no external forces are present. Choosing a generalized (with respect to the billiard walls) surface of section allows one to recover important aspects of the dynamics. These aspects can be classified according to the different angular velocity at which the surface of section is crossed, implied for each trajectory within the BKL maps.
The definition of quantum BKL probabilities for the first epoch of each era (\ref{12}) allows one to encode this information in the quantum description.\\
Long BKL eras have not been widely investigated yet in the quantum regime.\\
In \cite{berger}, a numerical Monte Carlo simulation of the behavior of the wavefunction of the BKL universe is performed, within the framework of Misner variables. Those regimes are analyzed, where the billiard ball is 'trapped in the minisuperspace potential'.  
The symmetries of the potential and those of the wavefunctions are compared. An effect is outlined, such that the features that cause classical chaos in the potential imply a correlation between a large value of the volume of the universe and a high value of the anisotropy.
The shape of the wavefunction is found to follow that of the potential close to the cosmological singularity and to fragment into fingers that point in the directions of the corners. In \cite{berger}, the investigation is motivated by a comparison with different studies, in which the wavefunction has been found to be large for a very small value of the anisotropy, as in \cite{hawking84}, \cite{amsterdamski85}, \cite{moss85}, \cite{furusawa86}, \cite{furusawa861}.\\
\\
A comparison of the Misner variables and of the Iwasawa variables on the unit circle has not been performed yet. The main difference between the two settings is the behavior of the potential.\\
Indeed, in the Mixmaster model, the potential is represented as a triangular well with moving walls, the walls moving 'outwards' with respect to the center of the potential, thus enlarging the space allowed for the motion of the billiard ball. In this case, the chaotic properties of the system are discussed according to the possibility for the billiard ball to reach the shifting walls at each bounce.\\
On the unit circle, the walls that delimit the triangular potential are steady.\\
It is easily verified that the two descriptions admit the same asymptotic (towards the cosmological singularity) limit on the UPHP, in the case of pure gravity in $4=3+1$ spacetime dimensions.\\
One of the advantages of the Mixmaster model is the possibility to identify the Hamiltonian variable which plays the role of Hamiltonian time as the isotropic growth of the universe, thus allowing one to interpret the two anisotropy variables as space variables.\\
In particular, a small value of the anisotropy is linked with a situation where no preferred oscillation corner is outlined. In other words, this corresponds to short trajectories taking place into different corners of the billiard.\\
A situation of a high value of the anisotropy corresponds to the physical configuration, for which there exists a preferred corner of oscillation.\\
This information can be linked to the present framework by considering that a sequence of short epochs takes place into different corners of the billiard, while a long era takes place into a single (preferred) corner of the billiard.\\
From the same point of view, a small value for the isotropic volume of the universe represents times close to the cosmological singularity, while a larger value of the isotropic volume of the universe corresponds to the opposite physical situation, as long as all the approximations implied in this solution hold.\\
\\
From the quantum point of view, no correlation has been clearly outlined yet, able to match the Misner hypothesis on high occupation numbers\cite{misn}, the quantum BKL numbers $n$ and the Fourier decomposition of the Maass wavefunctions found on the UPHP.\\
The statistical interpretation of the wavefunction according to the classical BKL trajectory, is based on the semiclassical interpretation of the wavefunction as evaluated on a classical trajectory characterized by the different classical symmetry-quotienting mechanisms.\\
In the statistical analysis accomplished in \cite{sinai83} and \cite{sinai85}, the properties of the BKL map are found to be such that, even though the occurrence short eras is mostly favored, given a long sequence of epochs, a randomly-chosen era within this sequence is most likely to contain a high number of epochs.\\
\\
In the present scheme, the situation described in \cite{berger} corresponds to the occurrence of a large value of $n$ in the first BKL era: in this case, indeed, the billiard ball is forced to perform a large number of oscillation between the same corner of the billiard potential.\\
A glance at long BKL eras in the quantum regime can help one shed light on these still unexplored features of the model.\\
Given an $n=1$ era and a long $n$ era, it is possible, by construction, to find a surface of section $u=u^*$, with $-1<u^*<0$, such that the first trajectory of each such BKL eras intersect. (In particular, for $n$ large, this value will tend to $-1$ from the right). This follows from the irregular tiling of the UPHP as far as the length of each BKL era is concerned.\\
An $n=1$-epoch era consists in only one oscillation on a given corner; a long era consists of several oscillations within the same corner. In the last case, however, the billiard ball approaches most to the vertex (as the vertex is placed on the absolute of the UPHP, which corresponds to the cosmological singularity) of the billiard only at (approximatively) ''half-way'' the era. The picture in which the billiard ball is close to one of the vertices of the billiard table corresponds indeed to (approximatively) half the total number of iterations of the epoch map.\\
Furthermore, as one of the advantages of the use of the BKL symmetry-quotienting of the dynamics, one can recover information on the content of epochs of each era by the knowledge of the value of the variable $u^+$.\\
For this, it is possible to study the proerities of the semiclassical limit of the wavefunction according to the continued-fraction decomposition of the BKL trajectory on which the semiclassical limit is performed. From the quantum point of view, this procedure allows one to define these trajectories without letting the dynamics evolve inside a corner.\\
The evaluation of the thickness of the slab $\sigma$ of the energy-shell reduced Liouville measure defined in Eq. (\ref{6}) for the two different eras takes into account also the value of the variable $u^-$. As a result, $\mid\cos \theta\mid$ tends to $1$ for long epochs, but much smaller than $1$ for short epochs: even though the initial trajectories cross at the point $u=u^*$, and are therefore defined on the same point of the billiard table on the UPHP, they cross the Poincar\'e surface of section at a different (angular) velocity.\\
The comparison of the angular velocities of the billiard ball crossing the surface of section at the first epoch of each era is most effectively achieved, on the UPHP, by considering the BKL epoch map or the CB-LKSKS (named after Chernoff- Barrow- Lifshitz- Khalatnikov- Sinai- Khanin- Shchur map) map, rather than the full unquotiented dynamics, as both maps are defined according to the value of the variable $u^+$ at the first epoch of each era.\\
From the quantum point of view, the semiclassical limit of the wavefunction is obtained by considering the wavefunction as evaluated on these trajectories, as the wavefunction depends only on $u$ and $v$. Even though their value is therefore the same at the same point on the Poincar\'e surface of section $u=u^*$, from the quantum point of view, the definition of quantum BKL probabilities (\ref{12}) allows one to characterize the wavefunction according to the different angular velocity at which the Poincar\'e surface of section is crossed.\\
\\
The comparison of the probabilities ${\bf P}_{n, s}$, Eq. (\ref{12}), and ${\bf P}_{n', s}$ with the properties of the BKL map and of the CB-LKSKS map within this framework is understood by the characterization of the quantum BKL probabilities as describing the semiclassical limit of the probability for a wavefunction to be evaluated on a given classical BKL trajectory: this characterization allows one to specify whether the trajectory is the first of an era, as in the case of ${\bf P}_{n, s}$, or the $n'$-the of an era.\\
Form the classical point of view, this difference is described by the statistical properties of the billiard maps, which imply the different symmetry-quotienting mechanisms.\\
From the quantum point of view, this properties allows one to evaluate the quantum probability associated to given initial conditions for the characterization of the cosmological singularity.\\
\\
Indeed, it is possible to specify these initial conditions according to the statistical properties of the BKL map and of the CB-LKSKS map for the quantum version of the model by the integration domain of the variable $u^-$ in the quantum probabilities (\ref{12}) and (\ref{13}). The different domains allow one to encode in the quantum probability the notion of whether the epoch considered is the first one of an era, or a different one.\\
\\ 
The relevance of surfaces of section in the characterization of the quantum dynamics is then understood as due to the possibility to define the symmetry properties of the wavefunction without choosing any kind of boundary conditions for the quantum system, to connect the quantum states to a classical trajectory, and to encode in this specification to the statistical properties of the BKL maps and of the CB-LKSKS maps.\\
\section{Periodic trajectories and scars\label{section5a}}
The different symmetry-quotienting mechanisms imply different statistical properties as far as the analysis of the length of the orbits is concerned. Moreover, the full unprojected dynamics has different statistical properties with respect to the projected one.\\
\\
The statistical properties of the probability distribution (\ref{5}) obtained from (\ref{symp}) imply that $n$-epoch orbits are more probable for small $n$. A two-variable map of the unit square onto the unit square, for suitable functions of the variables $u$ and $v$ of the UPHP, reveals that, even though the most probable BKL era consists of $1$ epoch, nevertheless, if a long sequence of epochs is considered, an epoch is most probably belonging to a large-$n$ era. The implications of this property have not been fully clarified yet, as far as the full billiard dynamics is concerned.\\
It is crucial to recall that the statistical analyses available at the moment concern only the properties of a finite sequence of epochs.\\
The full complete statistical description able to describe the precise properties of each sequence of epochs obtained from the continued-fraction decomposition of the statistical variables, which faithfully reproduce the initial values of the statistical maps, is still a tantalizing investigation theme, for both the occurrences of a finite continued-fraction decomposition and that of an infinite one. The interest in these kinds of approaches is due to the correspondence between cosmological billiards and the precise characterization of the initial values of the variables which determine the properties of the most general solution to the Einstein field equations is concerned, as it, on its turn, determines the properties which are exhibited by the present Universe, as far as the $4=3+1$ dimensional case is concerned. On the other hand, in the case of an infinite continued-fraction decomposition for the variable $u$, the occurrence of periodic continued fractions has to be considered as a special case.\\
\\ 
\subsection{Short eras}
Several studies have been proposed, in which the properties of a short sequence of epochs are mimicked by less complicated models, from which it is easier to extract information about some particular features of the dynamics. Far form reproducing the exact behavior of cosmological billiards, these models nevertheless compare the statistical properties of better-analyzable objects to those of the models obtained form the cosmological point of view, both at the classical level and at the quantum one.\\
\\
In this direction, a very important study has been proposed in \cite{cornish}, as far as the classical properties of the one-variable BKL maps are concerned, when only a short sequence of epochs is considered.\\
The system analyzed in \cite{cornish}, Eq. (3.15) is a two-variable map for the variables \sffamily{u} \rmfamily and \sffamily{v} \rmfamily , consisting of a symmetry-quotiented map for the variable \sffamily{u} \rmfamily (corresponding to the $u_{\rm{BKL}>1}$ definition of \cite{Damour:2010sz}) and its inverse for the variable \sffamily{v} \rmfamily .\\
As, in the sense specified in the above, the most frequent eras are those consisting of a small number of epochs, a phenomenological modification to the map has been proposed, according to which a maximum value of epochs (which, for future purposes, is defined as) $N$ is imposed (and which corresponds to $N=k-1$ for the definition of the parameter $k$ in the analysis of the map Eq. (3.15) in \cite{cornish}). Consequently, the value $N$ corresponds to the maximum length of a BKL era: for longer eras, corresponding to a higher value of the variable \sffamily{v} \rmfamily , the map imposes that this trajectory be neglected.\\
The only kind of never-ending oscillating trajectories are selected as only those whose initial values correspond to a periodic continued-fraction expansion, of which each digit represents only a number $n\le N$ of epochs, and singular trajectories are automatically avoided. This result is achieved by the implementation of a kind of Farey map.\\
\\
In \cite{levin2000}, the chaotic motion of a billiard ball in a regular octagon has been analyzed, and several important features have been pointed out.\\
A map has been proposed, for which all the sides of the octagon are identified. This identification corresponds to the definition of a BKL era map.\\
The presence of periodic orbits has been considered both at the classical level and at the quantum one.\\
In the quantum regime, several investigations of different models have outlined the presence of 'scars' in the wavefunction. Scars of the wavefunctions are described as the phenomena for which the absolute value of the wavefuntion corresponding to the classically-chaotic motion of a billiard ball is enhanced on the values of the variables that classically correspond to periodic configurations. In particular, wavefunctions are observed to assume as the most probable semiclassical trajectories the configurations corresponding to periodic oscillations; any deviation from a periodic one is observed to be most probable as a change to a different periodic configuration. Short-period periodic orbits are most favored.\\
\\
In \cite{cornish}, the features of the version of the Farey map which is used for the analysis of short-period periodic orbits are such that the fractal structure defined by the set of trajectories invariant under the billiard map can be most easily investigated.\\  
\\
The description of scars as unexpected thickenings of the probability density for a wavefunction to be located in correspondence of unstable periodic trajectories was outlined by \cite{heller} for the case of stadium billiards.\\
In \cite{bogomolny}, the presence of (a particular kind of) scars is investigated for plane polygonal billiards, and the connection between the energy level of a periodic trajectory and the corresponding classical periodic configuration for the presence of a scar in the quantum model is outlined.\\
Within this framework, scars for the wavefunctions are interpreted in \cite{levin2000} as due to the presence of an underlying structure in the quantum regime corresponding to the classical fractal structure.\\ 
The properties of scars in the wavefunction for quantum systems, whose classical behavior is that of a billiard, but which is defined on a different geometry, are connected with the quantum maps and their semiclassical limit in \cite{giller} and \cite{kelmer}.\\
The characterization of the semiclassical limit of the wavefunction for cosmological billiards obtained within the present allows one to gain information about the definition of a quantum state corresponding to a classical BKL configuration: the invariance of the wavefunction under the transformation that define the symmetry-quotiented maps is crucial in this description, and the consideration of a Poincar\'{e} surface of section allows one to implement the procedure without specifying the kinds of boundary conditions imposed on the wavefunction. On the contrary, in \cite{bogomolny}, the kind of boundary conditions obeyed by the wavefunction have to be specified.\\
\\
These phenomena have not been investigated yet for the geometry characterizing cosmological billiards. The role of the classical statistical distribution of periodic orbits within this framework has not been pointed out yet. In the simplest case, i.e. for the case of the one variable era map, periodic orbits are represented, in the restricted phase space, by the corresponding value of the $u^+$ variable, while the variable $u^-$ is allowed to range within the complete interval in which it is defined, i.e. $-2< u^-<-1$. For, this, the definition of the quantum probabilities as in (\ref{12}) is not of direct application.\\
Nonetheless, the role of periodic orbits can be schematized by following a slightly different procedure, which is aimed at restriciting the set of all possible initial values for the variable $u^+$ to those such that the highest number of epochs contained in an era is $N$, and then at focusing the attention on the 'Silver Means' corresponding to such set of possible values. 'Silver Means' are defined as a generalization of the 'Golden Ratio', i.e. those numbers whose continued fraction expansion is given by $[n-1; n, n, ...]$: within this framework, such a value for $u^+$ (and of the corresponding one for $u^-$) is the most direct example of periodic trajectory. According to this viewpoint, the comparison with the standard expression for the BKL probabilities (\ref{5}) is the most straightforward.\\
The implementation of the Farey map on the values of the restricted phase space of the billiard system corresponding that hypothesized for the Farey map Eq. (3.15) in \cite{cornish} leads to the selection of particular periodic BKL trajectories. Non-periodic trajectories whose value of the variable $u^+$ consists of digits which correspond to $n<N$-epoch are not considered; all such trajectories are represented, in the restricted phase space, by values of the variable $u^+$ which differ from the 'Silver Mean' for a small quantity.\\
The implementation of the standard BKL map on the restricted phase space corresponding to the prescription of the Farey map Eq. (3.15) in \cite{cornish} is not fully consistent, because some trajectories would be mapped to regions of the restricted phase space, which are considered as available for the system for a BKL map but not for the Farey map. The inequivalence between the two maps is further discussed in \cite{cornish}.\\
On the other hand, the regions of the phase space corresponding to periodic orbits are a set of disconnected points, on which the definition of probabilities as the (according to the measure (\ref{symp}))oriented areas is not of direct generalization.\\
In \cite{isola}, a comparison between the Gauss map and (a version of) the Farey map has been accompliched. Nevertheless, the techniques used in \cite{isola} do not apply to any version of the BKL maps or of the CB-LKSKS maps needed to reproduce such a set of disconnected points on a suitable portion of the reduced phase space.\\
\\
The interest in considering only a portion of the restricted phase space is motivated, from the physical point of view, by the fact that any modification of the BKL maps corresponds to a modification of the Einstein field equations: this can be due either to a different characterization of space-time points in the asymptotic limit close to the singularity, or to the consideration of some kinds of matter.\\
\\ 
If one has to specify this scheme for the present model, it is interesting to consider a restricted phase space such that the variable $u^-$ still ranges in the interval defined for the original two-variable BKL map, $-2\ge u^-\ge -1$, while the $u^+$ variable is defined on $N$ disconnected intervals, whose definition has the same effects of the hypothesis of a  phenomenologically-truncated map. One obtains
\begin{subequations}\label{17}
\begin{align}
&n-1+1/N\ge u^+ \ge n,\ \ n\ge N\label{17a}\\ 
&-2\ge u^-\ge -1 \label{17b}
\end{align}
\end{subequations}
\\
According to the properties of the continued-fraction decomposition of the variable $u^+$, this choice corresponds to the regions of the phase space needed to evaluate the probability for both the first era of a sequence and the second one to contain a number of epochs smaller than $N$. This simplified scheme is effective in describing these probabilities for the first two eras, i.e. at the first iteration of the CB-LKSKS map. Any further specification would bring further prescriptions on the portion of the restricted phase space available for the dynamics without adding any precise information about purely periodic trajectories. 
The regions of the restricted phase space, modified according to (\ref{17}), are $N$ disconnected 'rectangular' boxes. The shape of each rectangular box corresponds to a set of available values for the variable $u^+$.\\
\\
This choice (\ref{17}) corresponds to phenomenologically modify the restricted phase space, which allows for a very precise comparison of the physical properties of cosmological billiards and the mathematical ones, and on which the BKL maps and the CB-LKSKS maps act. Nonetheless, this choice does not correspond to any precise modification of the restricted phase space, as the variable $u^-$ is not considered.\\
In particular, the choice to apply a map for the variable $\textit{v}$ in \cite{cornish} does not affect the definition of the variable $u^-$, as the two variables encode different kinds of information, as far as the dynamics of the primordial universe is concerned.\\
\\
Each rectangular box defined in the restricted phase space (\ref{17}) automatically contains the value value of $u^+$, for which an $n$ periodic era takes place. These periodic configurations are the simplest example of periodic orbits, which generalize the notion of 'Golden Ratio' to that of 'Silver Means'.\\
 For the choice $-2\ge u^-\ge -1$, all kinds of periodicities are possible, according to the different symmetry-quotienting mechanisms. 
\\
As a result, the probability $P_{n,N}$ for an era to contain $n<N$ epochs and to be followed by another era of maximum length $N$ is obtained as the oriented area (according to the non trivial measure (\ref{symp})) of the pertinent region of the restricted phase space $A_{n,N}$, normalized by the total area $A_N$ of the restricted phase space available for this description.\\
The area $A_{n,N}$ is evaluated as
\begin{equation}
A_{n,N}=\int^{-1}_{-2}\int^{n}_{n-1+\tfrac{1}{N}}du^+=\tfrac{1}{2}\ln\tfrac{(n+1)(n+1+\tfrac{1}{N})}{(n+\tfrac{1}{N})(n+2)},
\end{equation}
and the total area $A_N$ of the restricted phase space available for the oscillatory phenomena is obtained as the sum of the areas of all the disconnected domains, i.e.
\begin{equation}
A_N=\sum_{n=1}^{n=N}A_{n,N}=\tfrac{1}{2}\ln 2\tfrac{N+1+\tfrac{1}{N}}{(N+\tfrac{1}{N})(N+2)}.
\end{equation}
The probability for an $n\le N$-epoch era to take place, and to be followed by an era of maximum length $N$, with $n\le N$, is therefore
\begin{equation}\label{pnn}
P_{n,N}=\tfrac{\ln\tfrac{(n+1)(n+1+\tfrac{1}{N})}{(n+\tfrac{1}{N})(n+2)}}{\ln 2\tfrac{N+1+\tfrac{1}{N}}{(N+\tfrac{1}{N})(N+2)}}. 
\end{equation}
The features of these new modified probabilities $P_{n,N}$ in (\ref{pnn}) encode those hypothesized in \cite{cornish}, Eq. (3.15), as far as the variable $u^+$ is concerned.\\
The new expressions for the probabilities $P_{n,N}$ differ from the original ones, and, in addition, two particular regimes can be studied.\\
In fact, the interest of the present analysis is to compare the effects of the occurrence of short BKL eras and those of the occurrence of long BKL eras. The length of the modified eras has now to be compared with the maximum value $N$ chosen for the representation. Anyhow, it is reasonable to assume, as in \cite{cornish}, Eq. (3.15), that the maximum value $N$ be not to small, with respect to the predictions of the original BKL statistics.\\
In the limit $n<<N$, one obtains
\begin{equation}\label{pnn1}
P_{n,N}\simeq\tfrac{1}{\ln2}\ln\tfrac{(n+1)^2}{n(n+2)}\left(1-\tfrac{1}{Nn^2+(N+1)n+1}+\tfrac{2}{\ln2}\tfrac{1}{N}\right)
\end{equation}
In the limit of large $n$, i.e. in the limit where the length of the BKL era is comparable with that of the maximum value $N$, one obtains
\begin{equation}\label{pnn2}
P_{n,N}\simeq \tfrac{1}{\ln2}\tfrac{1}{n^2}\left(1+\tfrac{1}{N}\tfrac{2}{\ln2}\right).
\end{equation}
The expansions (\ref{pnn1}) and (\ref{pnn2}) obviously take into account all the orders relevant in the comparison with Eq. (\ref{5}) and its expansion for large $n$.\\
In particular, the expression obtained in (\ref{pnn1}) for small values of $n$ has to be considered as exact with respect to $n$, and as an asymptotic formula for $N$.\\
On the contrary, the expression in (\ref{pnn2}) has to be considered as a limiting one in the case of  both the two variables $n$ and $N$.\\
\\
\subsection{Periodic orbits and scars}
One of the aims of \cite{cornish} is to study the properties of periodic configurations of the one-variable epoch map and of the one-variable era map, by modifying the range of the $u^+$ variable, and by considering only purely periodic trajectories. Indeed, the procedure discussed in \cite{cornish}, as far as Farey trees are concerned, is aimed at isolating the points that correspond to periodic trajectories, consisting of a given sequence of $M$ eras, each of which contains a number $n_i\le N$ of epochs, for each $i\le M$. This strategy is equivalent at isolating the corresponding values of the variable $u^+$ in the restricted phase space.\\
Consequently, it is possible to argue that, for each allowed value of $n$, range of the variable $u^+$ is automatically restricted to a smaller range, in which the periodic configuration corresponding to a 'Silver Mean' is present. This modification automatically excludes the values of $u^+$, which are, in some extent, much different from that necessary for a periodic configuration.\\
As a result, the occurrence of a periodic orbit for each value of $n\le N$ is not only ensured, but also favored, both at the classical level and in the quantum regime.\\
\\
The modified model here presented, within this framework, selects disconnected regions of the restricted phase space, which automatically contain the corresponding 'Silver Mean'. In this case, the properties of the simplest periodic trajectories are examined by considering those regions of the restricted phase space, which contain the corresponding values of $u^+$.\\
\\
The expression for the probabilities (\ref{pnn}), also in the two different regimes (\ref{pnn1}) and (\ref{pnn2}) implies that, also when only short eras are taken into account, is still a decreasing monotonic function of the number $n$, for which short eras are favored also when only particular regions of the restricted phase space are taken into account.\\
\\
From the quantum point of view, it is straightforward to verify that the consideration developped for the expressions (\ref{a}) and (\ref{b}), which define the probability for a wavefunction to be evaluated on a given classical trajectory characterized according the definition of epochs and eras, are not changed if only the regions (\ref{17}) of the restricted phase space are taken into account. In fact, the behavior observed for (\ref{a}) is still valid when the modified probability (\ref{pnn2}) is taken into account, as the limit $P_{n, N}/P_{n+1, N}$ is not affected by any modification term at the order $\mathcal{O}(1/n)$.\\
Furthermore, the property of (\ref{b}) of being extremized by $n=1$ is not modified by the consideration of the disconnected regions (\ref{17}) of the restricted phase space.\\
\\
Specifying the integration domains of the expressions of the quantum probabilities ${\bf P}_{n, s}$ (\ref{12}) according to the modified range (\ref{17}) of the restricted phase space leads to the expression for the probability ${\bf P}_{n, N, s}$ of a wavefunction belonging to the energy-shell $s$ to be evaluated on a classical trajectory corresponding to an era containing $n<N$ epochs, and to be followed by a new BKL era containing a number $n<N$ epochs
\begin{equation}\label{18}
{\bf P}_{n, N, s}=\int_{-2}^{-1}du^-\int_{n-1+\tfrac{1}{N}}^{n} du^+\int_{\theta_2}^{\theta_1}\tfrac{d\theta}{\sin\theta}\tfrac{1}{(u^+-u^-)^2}\mid\Phi_{s}(u^-,u^+,u^*)\mid^2,
\end{equation}
\\
The same modification of the integration domains on the restricted phase space for the probability ${\bf P}_{n', s}$ of a wavefunction of the energy-shell $s$ to be evaluated at the $n'<N'<N$  BKL epoch of an $N'$-epoch BKL era and to be followed by a new BKL epoch containing a maximum number $N$ of epochs leads to the expression  
\begin{equation}\label{19}
{\bf P}_{n', N, s}=\int_{-2-n'}^{-1-n'}du^-\int_{N'-n'+\tfrac{1}{N}}^{N'-n'} du^+\int_{\theta_2}^{\theta_1}\tfrac{d\theta}{\sin\theta}\tfrac{1}{(u^+-u^-)^2}\mid\Phi_{s}(u^-,u^+,u^*)\mid^2.
\end{equation}
From the probabilities ${\bf P}_{n, N, s}$ and ${\bf P}_{n', N, s}$, Eq.'s (\ref{18}) and (\ref{19}), respectively, one learns that the choice of any modification of the restricted phase space does not modify, and is not modified by, the consideration of any particular surface of section $u=u^*$. Indeed, each trajectory necessarily intersects any such surface of section.\\
\\
Even though, in the expressions of ${\bf P}_{n, N, s}$ and ${\bf P}_{n', N, s}$, no evidence for the appearance of scars in the wavefunction is directly worked out, it is possible to infer that, also in the quantum regime, the ranges of the variables $u^+$ and $u^-$ are modified in a way such that the occurrence of short period periodic orbits is favored. Scars in the wavefunctions can therefore be related to the results of the present investigation by considering the the regions of the allowed phase space on which the wavefunction is mostly probably found are restricted along those values of the variable $u^+$, which correspond to a periodic orbit.\\
The role of the symmetry-quotienting mechanisms implied in the BKL description of the dynamics, as well as in the CB-LKSKS map, is encoded in the characterization of the probability for a quantum configuration by the angular velocity at which the Poincar\'{e} surface of section is crossed.\\
This feature is not observed in the probabilities ${\bf P}_{n, s}$ and ${\bf P}_{n', s}$, given in Eq.'s (\ref{12}) and (\ref{13}), respectively, and constitutes a step forward in the definition of scars in the semiclassical limit of the wavefunctions for cosmological billiards in $4=3+1$ space-time dimensions.\\
\\
From the physical point of view, these properties are related to the initial conditions which characterize the solution to the Einstein field equation within the asymptotical (towards the singularity) limit of the Bianchi IX dynamics within the framework of the BKL paradigm. From this characterization, one learns that short eras are most probable, as far as the initial conditions are concerned, also within the framework of the modification (\ref{17}).\\
This way, also within this framework, short eras are the most probable characterization of the initial conditions for the values of $u^+$, which, on its turn, characterizes the physical features of the cosmological singularity.\\   
 
\section{Outlook\label{section5}}
In this analysis, some features of the cosmological billiards describing pure gravity in $4=3+1$ dimensions have been investigated. Particular attention has been devoted to the difference between the geometrical properties of the billiard domains and the dynamical properties of the billiard chaotic evolution. In fact, when symmetry quotienting mechanisms have to be implemented, such that BKL maps can established, some differences between the two procedures arise, and they can be interpreted, at least in the most straightforward analysis, according to the content of Weyl reflections in these maps: even though it is, in principle, possible to identify the sides of the billiard without Weyl reflections, the information needed to consistently encode the dynamics requires this kind of reflections.\\
These only apparent 'mismatch' is due to the features of the Einstein field equations, where the $3$ Kasner directions are considered as an unordered set, while the symmetry-quotienting mechanisms are based on the orientation of the geodesics.\\
A possible strategy for reconciling these different viewpoints is to consider the $3$-dimensional energy-shell reduced Liouville measure $\Omega^{(3)}_L$ in (\ref{omega3}), by which the content of Weyl reflections in the UPHP can be taken care of.\\ 
\\
From the quantum point of view, the difference between imposing boundary conditions for the wavefunction according to the features of the billiard table and to those of the BKL maps for the dynamics has been investigated in \cite{Lecian:2013cxa}, and the definition of quantum BKL probabilities has been formally proposed. This definition can be made more effective, such that more physical information can be extracted from it, if the wavefunction is expressed in the $(u^-,u^+)$ variables of the restricted phase space by introducing a suitable Poincar\'e section on the UPHP, and then by formally integrate it by means of the invariant measure $\Omega^{(3)}_L$. By this procedure, it has been possible to investigate what quantum BKL numbers are probabilistically favored. Interpreting the available information according to the continued-fraction decomposition of the $(u^-,u^+)$ variables, is is possible to infer that $1$ epoch BKL eras are the most probable also in the quantum regime.\\
\\
Because Poincar\'e sections of the big billiard (i.e. on the UPHP) can be verified to be crossed, by construction, by different kinds of trajectories corresponding to the different kinds of eras (i.e. not only of the $ba$ type), the action of \textit{any} symmetry-quotienting mechanism requires further specifications, as it has been necessary to impose throughout the present analysis.\\
From the quantum point of view, these procedures allow one to gain physical insight about the BKL configurations describing the succession of epochs and eras without imposing any specific boundary conditions. Indeed, the statistical meaning of BKL maps can be researched for in the Poincar\'e return map of the bounces of the billiard ball on the sides of the billiard table, on which boundary conditions are usually imposed boundary conditions on.\\
\\
\subsection{Properties of the wavefunction}
The role of the Poincar\'{e} surfaces of section in defining the properties of a BKL trajectory according to the angular velocity at which the billiard ball crosses the surface of section and to relate them with the epoch maps and to the era maps has been pointed out.\\
This definition applies to the comparison of several phenomena, which take place on the billiard table, which had not been characterized from this point of view yet.\\
The angular velocity which characterizes the classical geodesic path followed during the free-flight evolution of a Kasner epoch allows to compare the first epoch of two different (i.e. corresponding to a different content of BKL epochs) BKL eras.\\
This mechanism applies also for the characterization of an epoch, which is not the first of a given BKL era.\\
From the classical point of view, this new characterization can be employed for the comparison of the statistical properties of the two-variable CB-LKSKS map defined for the two corresponding unit-square variables as a map of the unit square onto the unit square, for which it is possible to appreciate the phenomenon according to which, if a long sequence of epochs is taken into account, it is most probable that a randomly-chosen epoch belongs to a long BKL era.\\
\\
The comparison of the known results about scars in the wavefuntion and the properties of the symmetry-quotienting of the dynamics which characterizes the BKL paradigm and the CB-LKSKS map have been here examined by considering only particular regions of the restricted phase space, which correspond to an initial sequence of two short eras.\\
\\
The comparison of the appearance of scars in the wavefunction corresponding to the semiclassical limit of the quantum regime of cosmological billiards in $4=3+1$ spacetime dimensions and the characterization of the quantum BKL probabilities by means of the corresponding regions of the restricted phase space is a new interesting feature, which, to the presence advance of the literature, had not been pointed out yet.\\
\\
The role of surfaces of section in the definition of the billiard dynamics from this point of view is then understood to be that of taking into account a symmetry-quotienting mechanism without specifying any boundary condition on the wavefunction, ad to characterize the semiclassical limit of the wavefunctions by the correspondence with classical BKL trajectories, for which a precise statistical analysis is available.\\
\\
The importance of Poincar\'{e} surfaces of section in the attempt of a characterization of scars in the wavefunction of cosmological billiards has therefore been investigated within this framework.\\
Astonishingly, the wavefunction, in its semiclassical characterization, according to the properties of the statistics of the BKL map and of the CB-LKSKS map, is found to be most probably located on the regions of the restricted phase space containing the BKL trajectories which correspond to the 'Silver Means', of which, those corresponding to the lowest values of $n$ are those for which scars of the wavefunctions are usually provoked. 

\section{Concluding remarks\label{section6}}
In this paper, some particular aspects of the cosmological billiards describing pure gravity in $4=3+1$ space time dimensions have been investigated, and some light has been shed both at the classical level and as far as the quantum regime is concerned. In particular, the phenomena have been analyzed, for which the content of Weyl reflections on the maps on the UPHP requires the introduction of a $3$-dimensional invariant measure, when physical objects have to be defined by means of integration of the corresponding invariants on suitable regions of the phase space.\\
From the classical point of view, the implementation of BKL maps on for the motion inside the UH has been considered as far as the action of BKL in exchanging the role of the Kasner directions is concerned.\\
Form the quantum point of view, the definition of quantum BKL probabilities for the BKL quantum numbers contained in the wavefunctions solving the eigenvalue problem on the UPHP has been refined by the formal result (\ref{12}). In this case, the squared modulus of the wavefunction has to be integrated by means of the $3$-dimensional energy-shell restricted Liouville measure.\\
Furthermore, by analyzing the behavior of the wavefunction via the tools refined in \cite{Damour:2010sz} and \cite{Lecian:2013cxa}, by the comparison of the derivatives of the wavefunctions with respect to the value of the oriented endpoints which encode the the succession of BKL epochs in the dynamics, the most probable configuration form the quantum point of view has been defined to the still that of a sequence of $1$-epoch BKL eras. In particular, this can be achieved by the periodic configuration defined by the values $u^-=-\Phi$ and $u^+=\phi$, or by a small perturbation of it.\\
More in detail, it is possible to infer this information by interpreting the values of the oriented endpoints $u^+$ and $u^-$ as encoding the number of BKL epochs in the future evolution and in the past one, according to the continued fraction decomposition of these parameters, respectively.\\
\\
The characterization of the dynamics of cosmological billiards, obtained for those phenomena which find a suitable characterization by the Poincar\'{e} surfaces of section in the classical regime, allow one to grasp information about the semiclassical limit of the quantum regime. For the case of cosmological billiards, the properties which define the statistical description of the dynamics within the framework of the different symmetry-quotienting mechanisms are found to imply, at the quantum level, that wavefunctions are mostly probably evaluated for those regions of the restricted phase space, which contain short-period periodic orbits: these kinds of orbits are those which correspond the the phenomenon of scars in the wavefunction of the quantum version of several kinds of billiards.\\ 
Differently from the other approaches to this phenomenon, for this case it is possible to evaluate also the contribution of short eras, which are not part of any periodic sequence. 
\\
The paper has been organized as follows.\\ 
In Section \ref{section2}, the features of cosmological billiards and their description by means of suitable symmetry-quotienting maps (the BKL maps) have been recalled.\\
In Section \ref{section3}, the physical problems for which a $2$-dimensional scenario requires a $3$-dimensional phase space have been outlined. The use of Poincar\'e sections in the definition of Poincar\'e return maps has been adopted for the description of particular features of cosmological billiards.\\
In Section \ref{section4}, the analysis of quantum BKL numbers has been refined according to these tools, and the formal definition of quantum BKL probabilities has been rested according to the $3$-dimensional energy-shell reduced Liouville measure $\Omega^{(3)}_L$. As a result, the most probable configuration on which the wavefunction has to be calculated is one close given by the periodic succession of $1$-epoch eras, i.e. for $n=1$ also form the quantum point of view.\\
In Section \ref{section5a}, the role of short-period periodic orbits has been discussed, and the properties of the simplest periodic configurations are outlined, both for the classical scheme and in the quantum version of the model. Accordingly, the connection between this characterization of the dynamics and the occurrence of scars for the wavefunctions of cosmological billiards has been pointed out\\.
In Section \ref{section5}, the role of the invariant measure $\Omega^{(3)}_L$ in reconciling geometrical features of the billiard tables with that of the dynamics, which is encoded by a suitable composition of Weyl reflections, has been clarified: the connection between $(i)$ the role of the Poincar\'{e} surfaces of section, $(ii)$ the characterization of the wavefunction of cosmological billiards according to the angular velocity at which the surface of section is crossed by the billiard ball in the classical description, $(iii)$ the statistical features of the different maps, for which the different symmetry-quotienting mechanisms are obtained, $(iv)$ the characterization of the wavefunctions for cosmological billiards according to the classical trajectory on which it is evaluated in the semiclassical limit, $(v)$ the definition of quantum probabilities for the quantum configurations which correspond to the characterization of the dynamics in the BKL paradigm and $(vi)$ the \textit{possibility} of the occurrence of scars in the wavefunction of cosmological billiards have been provided with a rigorous connection.
\section*{Acknowledgments}
OML is grateful to the Referee for having proposed the investigation themes followed in Section IVC and in Section V .

\end{document}